\documentclass[12pt]{article}
\usepackage{epsf}
\usepackage{amsmath}
\usepackage{graphics}
\usepackage{cite}

\renewcommand{\thefootnote}{\#\arabic{footnote}}
\begin{document}

\newcommand{\gtrsim}{ \mathop{}_{\textstyle \sim}^{\textstyle >} }
\newcommand{\lesssim}{ \mathop{}_{\textstyle \sim}^{\textstyle <} }

\newcommand{\rem}[1]{{\bf #1}}

\renewcommand{\thefootnote}{\fnsymbol{footnote}}
\setcounter{footnote}{0}
\begin{titlepage}

\def\thefootnote{\fnsymbol{footnote}}

\hfill arXiv:13mm.nnnn [hep-ph]

\begin{center}

\bigskip
\bigskip
{\Large \bf Higgs couplings to $t$,$b$, and $\tau$ with Flavor Symmetry}

\vspace{1.0in}

{\bf Paul H. Frampton\footnote{frampton@physics.unc.edu}}

\vspace{1.0in}

{Department of Physics and Astronomy, UNC-Chapel Hill, NC 27599.}

\end{center}

\vspace{1.0in}

\begin{abstract}
Experimantal study at LHC of the possible Higgs boson
should soon provide accurate estimates of the Yukawa couplings
$Y_{H\bar{f}f}$ for $f=t,b,\tau$. In the presence of a
non-abelian discrete flavor symmetry $G_F$ ({\it e.g.} $T^{\prime}$)
the usual prediction that $Y_{H\bar{f}f} \propto m_f$
should be inexact, and departures therefrom will provide
through $G_F$ valuable input to an improved
derivation of the quark and lepton mixing matrices.
\end{abstract}

\end{titlepage}

\renewcommand{\thepage}{\arabic{page}}
\setcounter{page}{1}
\renewcommand{\thefootnote}{\#\arabic{footnote}}

The 2012 discovery of a resonance at $\sim 125$ GeV at the LHC
\cite{CMS,ATLAS}, strongly suggestive of the Higgs boson,
has naturally caused intense interest. It preliminary properties
are consistent within significant errors with the Higgs particle
predicted by the minimal standard model. 

The two-body decays which can be measured accurately
in the near future include $H\rightarrow\gamma\gamma$,
$H\rightarrow\bar{b}b$, and $H\rightarrow\bar{\tau}\tau$.
These are respectively sensitive to the Yukawa couplings
$Y_{H\bar{t}t}$ (through the top triangle contribution
which competes with the $W$-loop), $Y_{H\bar{b}b}$,
and $Y_{H\bar{\tau}\tau}$.

In the minimal standard model, the Yukawa couplings
$Y_{H\bar{f}f}$ appear in the simple form

\begin{equation}
Y_{H\bar{f}f} \bar{f}f H
\label{Yukawa}
\end{equation}
so that they are proportional to the masses

\begin{equation}
Y_{H\bar{f}f} \propto m_f 
\label{proportional}
\end{equation}
with proportionality constant $<H>^{-1}$ where $<H>$ is 
the vacuum expectation value of the Higgs field.

\bigskip

In a renormalizable model with a non-trivial flavor symmetry $G_F$,
which we will take here to be non-abelian and discrete, there must
be several Higgs and the Yukawa couplings of the lightest one $H$
will generally deviate from the simple proportionality
of Eq.(\ref{proportional}). Such deviations may likely be small
but crucial to understanding how the group $G_F$ operates. One
may even say that if the conventional prediction of
Eq. (\ref{proportional}) would hold exactly at high precision
then renormalizable $G_F$ models would be disfavored.

\bigskip

These statements are true for general renormalizable $G_F$ models.
To illustrate them, we focus on the choice $G_F = T^{\prime}$
\cite{FK1} and the minimal model discussed in \cite{FKM}.
The flavor group is $(T^{\prime} \times Z_2)$, and we shall
concentrate only on the third-generation couplings
$Y_{H\bar{f}f}$ for $f=t,b,\tau$.

\bigskip

The leptons are assigned 
under ($T^{\prime} \times Z_2$) as

\begin{equation}
\begin{array}{ccc}
\left. \begin{array}{c}
\left( \begin{array}{c} \nu_{\tau} \\ \tau^- \end{array} \right)_{L} \\
\left( \begin{array}{c} \nu_{\mu} \\ \mu^- \end{array} \right)_{L} \\
\left( \begin{array}{c} \nu_e \\ e^- \end{array} \right)_{L} 
\end{array} \right\} 
L_L  (3, +1)  &
\begin{array}{c}
~ \tau^-_{R}~ (1_1,-1) \\
~ \mu^-_{R}~ (1_2,-1) \\
~ e^-_{R}~ (1_3,-1) \end{array}
&
\begin{array}{c}
~ N^{(1)}_{R} ~ (1_1,+1) \\
~ N^{(2)}_{R} ~ (1_2,+1) \\
~ N^{(3)}_{R} ~ (1_3,+1). \\ \end{array}
\end{array}
\end{equation}
Imposing strict renormalizability on the lepton lagrangian 
allows as nontrivial terms for the $\tau$ mass only 

\begin{equation}
Y_\tau \left( L_L \tau_R H'_3 \right)
+
{\rm h.c.}
\label{Taulepton}
\end{equation}
where $H'_3$ transforms as $H'_3 (3, -1)$.

\bigskip

The left-handed quark doublets \noindent $(t, b)_L, (c, d)_L, (u, d)_L$
are assigned under $(T^{'} \times Z_2)$ to

\begin{equation}
\begin{array}{cc}
\left( \begin{array}{c} t \\ b \end{array} \right)_{L}
~ {\cal Q}_L ~~~~~~~~~~~ ({\bf 1_1}, +1)   \\
\left. \begin{array}{c} \left( \begin{array}{c} c \\ s \end{array} \right)_{L}
\\
\left( \begin{array}{c} u \\ d \end{array} \right)_{L}  \end{array} \right\}
Q_L ~~~~~~~~ ({\bf 2_1}, +1)
\end{array}
\label{qL}
\end{equation}

\noindent and the six right-handed quarks as

\begin{equation}
\begin{array}{c}
t_{R} ~~~~~~~~~~~~~~ ({\bf 1_1}, +1)   \\
b_{R} ~~~~~~~~~~~~~~ ({\bf 1_2}, +1)  \\
\left. \begin{array}{c} c_{R} \\ u_{R} \end{array} \right\}
{\cal C}_R ~~~~~~~~ ({\bf 2_3}, -1)\\
\left. \begin{array}{c} s_{R} \\ d_{R} \end{array} \right\}
{\cal S}_R ~~~~~~~~ ({\bf 2_2}, +1)
\end{array}
\label{qR}
\end{equation}

\noindent We must two new scalars $H_{1_1} (1_1, +1)$ and
$H_{1_3} (1_3, +1)$ whose VEVs
\begin{equation}
<H_{1_1}> = m_t/Y_t ~~~~ <H_{1_3}> = m_b/Y_b
\label{H13VEV}
\end{equation}
provide the $(t, b)$ masses.
In particular, no  $T^{'}$ doublet
($2_1, 2_2, 2_3$) scalars have been added. This 
allows a non-zero value only for $\Theta_{12}$. The other angles  
vanish making the third family stable

\bigskip

The Yukawa couplings to the third family of quarks are contained in

\begin{eqnarray}
{\cal L}_Y^{(quarks)}
&=& Y_t ( \{{\cal Q}_L\}_{\bf 1_1}  \{t_R\}_{\bf 1_1} H_{\bf 1_1}) \nonumber \\
&&
+ Y_b (\{{\cal Q}_L\}_{\bf 1_1} \{b_R\}_{\bf 1_2} H_{\bf 1_3} ) \nonumber \\
&&
+ {\rm h.c.}
\label{Yquark} 
\end{eqnarray} 

The use of $T^{'}$ singlets and doublets
permits the third family to differ 
from the first two and thus make plausible the 
mass hierarchies $m_t \gg m_b$,  $m_b > m_{c,u}$ 
and $m_b > m_{s,d}$ as outlined in \cite{FK1}. 

Such a model leads to the formula\cite{FKM} for the Cabibbo angle
\begin{equation}
\tan 2\Theta_{12} = \left( \frac{\sqrt{2}}{3} \right)
\label{Cabibbo}
\end{equation}

\noindent or equivalently $\sin \Theta_{12} = 0.218..$
close to the experimental value
$\sin \Theta_{12} \simeq 0.227$.

\bigskip

It can also lead to the successful relationship
between neutrino micing angles $\theta_{ij}$
\begin{equation}
\theta_{13} = (\sqrt{2})^{-1} \left| \frac{\pi}{4} - \theta_{23} \right|
\label{neutrinomixing}
\end{equation}
which is also in excellent agreement with the latest experiments\cite{EF}.

\bigskip

In such a model, the lightest Higgs $H$ is a linear combination

\begin{equation}
H = a H_{1_1} + b H_{1_3} + c H_{3}^{\prime} + ...
\label{lighthiggs}
\end{equation}

\noindent and the consequent Yukawa couplings are

\begin{equation}
Y_{H\bar{t}t} = a^{-1} Y_t, ~~ Y_{H\bar{b}b} = b^{-1} Y_b, ~~ Y_{H\bar{\tau}\tau} = c^{-1} Y_{\tau}
\label{modeified}
\end{equation}

\bigskip

The VEV $<H>$ is shared between the $<H_{\alpha}>$ ($\alpha = 1_1, 1_3, 3',...)$
irreps of $T^{\prime}$ and there
is no reason to expect $a=b=c=... =1$ so that
the proprtionality of Eq.(\ref{proportional}) will be lost.
In fact, if Eq.(\ref{proportional}) remained exact, the only solution would
be a trivial one where all states transform as $1_1$ of $T^{\prime}$
and the $G_F$ is inapplicable. The successes in \cite{FKM} and \cite{EF} 
would, in such a case, be accidental.

\bigskip

On the other hand, if Eq.(\ref{proportional}) is inexact, the evaluations
of the coefficients $a,b,c,...$ can then be used to understand more
perspicuously the derivations of mixing angles for quarks and leptons
given respectively in \cite{FKM} and \cite{EF}, in a first
clear departure from the minimal standard model.

\newpage

\begin{center}
{\bf Acknowledgements}
\end{center}

\noindent This work should have been supported by U.S. Department of Energy grant number 
DE-FG02-06ER41418.


\vspace{1.5in}



\begin{thebibliography}{99}
\bibitem{CMS}
J. Incalada, talk given at CERN on July 4, 2012.
CMS Collaboration. CMS-PAS-HIG-12-020.
\bibitem{ATLAS}
F. Gionetti, talk given at CERN on July 4, 2012.
ATLAS Collaboration. ATLAS-CONF-2012-093.
\bibitem{FK1}
P.H. Frampton and T.W. Kephart, Int. J. Math. Phys. {\bf A10,}
4689 (1995).
{\tt hep-ph/9409330}.
\bibitem{FKM}
P.H. Frampton, T.W. Kephart, and S. Matsuzaki,
Phys. Rev. {\bf D78,} 073004 (2008).
{\tt arXiv:0807.4713[hep-ph]}
\bibitem{EF}
D.A.Eby and P.H. Frampton, Phys. Rev. {\bf D} (in press).
{\tt arXiv:1112.2675[hep-ph]}
\end{thebibliography}
\end{document}